\documentclass[twocolumn]{revtex4}
\usepackage{graphicx}
\usepackage{graphics}
\usepackage{amssymb}
\usepackage{epstopdf}
\usepackage{color}
\usepackage{subfigure}
\usepackage{bbm}
\usepackage{setspace}

\renewcommand{\d}{{\textrm{d}}}

\newcommand{\paper}{paper}

\begin{document}

\title{Bell-inequality test for spatial mode entanglement of a single massive particle}
\author{Libby Heaney$^{1,2}$ and Janet Anders$^3$}
\affiliation{$^1$The School of Physics and Astronomy, University of Leeds, Leeds LS2 9JT, United Kingdom.\\
$^2$Centre for Quantum Technologies, National University of Singapore, Singapore 117543, Singapore.\\
$^3$Department of Physics and Astronomy, University College London, London WC1E 6BT, United Kingdom}

\begin{abstract}

Experiments violating Bell inequalities have formed our belief that the world at its smallest is genuinely non-local. While many non-locality experiments use the first quantised picture, the physics of fields of indistinguishable particles is captured most conveniently by second quantisation. This implies the possibility of non-local correlations, such as entanglement, between modes of the field. In this \paper,\, we propose an experimental scheme that tests the theoretically predicted entanglement between modes in space occupied by \emph{massive} bosons.  A successful test of this scheme would  not only indicate that mode entanglement is as genuine as particle entanglement, despite the particle number superselection rule, but would also prove that this superselection rule can be overcome.
\end{abstract}

\maketitle

\section{Introduction}

 Presumably every undergraduate course on quantum physics starts with the problem of a single particle in a box. The model serves to introduce the notion of wavefunctions, energy eigenlevels and the superposition principle. Astonishingly, even the sophisticated concept of quantum entanglement can be introduced with the help of this simple system. 

The powerful correlations of entanglement are a fundamental feature of quantum mechanics and an important resource for quantum computing \cite{NielsenChuang}. Experimental demonstrations of  entanglement between the internal degrees of freedom, such as polarisation, of pairs of localised particles, such as photons, have been achieved \cite{Aspect82}.  Entanglement also manifests itself between (continuous) external degrees of freedom, such as the position of individual ions in a chain \cite{latticepapers}. However, in continuous systems of indistinguishable particles, entanglement may exist between second quantised field modes \cite{Vedral03}, rather than between the particles themselves. One important choice of modes are spatial modes, since correlations stretching over space lie at the heart of phase transition effects, such as Bose-Einstein condensation (BEC).  

A number of works  \cite{Simon02,Anders06, Koji06}  establish sufficient conditions for the existence of entanglement between spatial regions of massive bosonic fields.  Moreover, spatial entanglement has been linked to the appearance of spatial coherence in the BEC at low temperatures \cite{Heaney07, Goold:09}, indicated by a slowly decaying classical two-point correlation function  \cite{Kheruntsyan}.
Many of these indicators of entanglement are experimentally easy to measure, such as the temperature of a bosonic gas \cite{Anders06}. Yet any of these tests assumes a model of the system to begin with and the conclusion depends heavily on the validity of this model. An indisputable way to avoid this problem and experimentally confirm spatial entanglement is via a Bell-inequality test between two local parties, each having access to only one spatial region. Such a Bell-inequality test requires no {\it a priori} knowledge about the system and while assumptions are made to construct a sensible Bell-test, the actual clicking pattern in an experiment either shows non-classical correlations, i.e. entanglement, or not - whatever the theory might predict.

However, to perform a Bell test for mode entanglement, each party has to measure their part in (at least) two different measurement settings and it was until now unresolved how to measure modes of a massive bosonic field in any  way other than in the particle number basis. This is why it is disputed  \cite{Greenberger96} whether mode entanglement is as `genuine' as discrete spin entanglement or if it is just an artefact of quantum statistics. In this \paper\, we solve this problem by uniting a number of ingredients from recent advances in theory and experiment to establish an unambiguous experimental method to confirm spatial entanglement, should it `really' exist. 

We consider a standard setup similar to the one proposed by Tan {\it et al.} \cite{Tan91}, where a single photon prepared in a superposition state of being in one of two paths leads to mode entanglement.  Non-locality of a single photon has been confirmed experimentally in \cite{Hessmo04}. We extend the photonic  scheme so that, despite the existence of a superselection rule for mass,  a Bell-inequality can still be analyzed for entanglement arising \emph{naturally} from a single massive boson distributed over two spatial modes. In our scheme, each spatial mode is locally mixed with a reference state at an atomic beamsplitter. This operation is essential as it locally introduces additional particles to the system, thus circumventing the superselection rule and allowing measurements in bases other than particle number. After the beamsplitting operation, detectors measure the number of atoms in each of the output modes and the outcomes are analysed in a Bell-inequality. The results show that certain measurement settings reveal entanglement of the spatial modes.  A successful test of our scheme would unambiguously  indicate that entanglement between two modes is as genuine as entanglement between the degrees of freedom of two or more particles.

\section{ Entanglement between spatial modes} 

For any confining volume, the energy modes of a bosonic field labelled by $k$, can be `excited' by applying creation operators $\hat a^{\dag}_k$, where $[\hat{a}_k,\hat a^{\dag}_l]=\delta_{kl}$, on the vacuum state, $\hat a^{\dag}_k |vac \rangle = |0_1 0_2 ... 1_k 0_{k+1} ... \rangle$.  An excitation is called `a particle', which in this \paper\, is a massive boson with corresponding quantised energy $E_k = \hbar \omega_k$, where $\omega_k$ is the frequency of the $k$-th energy eigenmode. Instead of describing the system using its energy modes, one can use a different set of modes, such as the \emph{spatial modes}. The description of the system in space can be obtained by a transformation via the energy eigenfunctions, $ \hat a^{\dag}_k | vac \rangle = \int \d x \, \phi_k(x) \, \hat \psi^{\dag}(x) \,| vac \rangle$,  where $\phi_k(x)$ is the $k$-th energy eigenfunction and $\hat \psi^{\dag}(x)$ creates a particle at point $x$ in space. Populating an energy mode with one particle is therefore equivalent to populating all spatial modes, that is, all points in space, in a superposed manner. The occupation number of each spatial point, $x$, is $\hat n(x) = \hat \psi^{\dag}(x) \hat \psi (x)$ with eigenvalues, $n(x) = 0, 1, 2, ...$. 

\begin{figure}[t]
   \begin{center}
\includegraphics[width=0.4\textwidth]{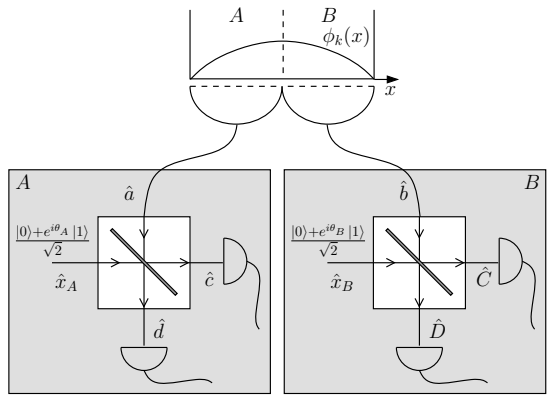}
       \caption{\label{fig:setup} A massive bosonic particle is distributed over a box, occupying, for example, the $k$-th energy eigenmode with spatial distribution, $\phi_k (x)$. Two parties each have access to a region of the box, $A$ ($B$), and coherently guide the boson from their region to a beamsplitter at which it meets a reference state, $|\theta^{A(B)}\rangle = \frac{1}{\sqrt{2}}(|0\rangle+e^{i\theta_{A(B)}}|1\rangle)$. Two detectors measure the number of bosons in the two output modes of the beamsplitter for each party. }
   \end{center}
\end{figure}

Let us divide space into two extended, non-overlapping regions, $A$ and $B$, see Fig.~\ref{fig:setup}, and populate them with a \emph{single} particle in the $k$-th energy eigenmode. The pure state of the two spatial modes is a superposition state
\begin{equation}\label{eq:purestate}
	|\psi \rangle = \hat a^{\dag}_k | vac \rangle = \alpha |01\rangle_{AB} + \beta |10 \rangle_{AB},
\end{equation}
where $|01\rangle_{AB}$ denotes a particle in region $B$ and no particle in $A$ and $\alpha$ and $\beta$ are normalised complex coefficients. For general values of the coefficients this state can \emph{not} be written as a product between the regions and is considered non-separable according to the mathematical definition of entanglement.   By demonstrating that mode entanglement can, in principle, violate a Bell inequality, we suggest that it is not just a mathematical feature of the quantum state, but could also be used as a quantum resource, see for instance \cite{Heaney:09}.

\section{ Bell-inequality for mode entanglement of massive particles }

Bell's inequalities \cite{Bell64} limit the maximum correlations that a bi-partite system can share under the assumptions of local realism.
In practice, one tests a variation of Bell's inequality proposed by Clauser-Horne-Shimony-Holt (CHSH) \cite{Clauser69} where two (arbitrarily separated) parties, $A$ and $B$, measure their share of a joint system locally, each in one of two settings.  The four joint expectation values of the measurements, $E(\theta^A_j,\theta^B_k)$, with $|E(\theta^A_j,\theta^B_k)| \le 1$ for $A$ measuring in setting $\theta^A_j$ and $B$ measuring in setting $\theta^B_k$ for $j, k =1, 2$, together form the CHSH inequality
\begin{equation} \label{eq:C}
	C = | E(\theta^A_1,\theta^B_1) +E(\theta^A_1,\theta^B_2) +E(\theta^A_2,\theta^B_1) - E(\theta^A_2,\theta^B_2) | \le 2,
\end{equation}
limiting classical correlations to a maximal value of 2.  All entangled quantum states in the Hilbert space $\mathcal{H}=\mathbbm{C}^2 \otimes \mathbbm{C}^2$ fail to obey at least one such inequality \cite{Horodecki95} and the existence of non-local correlations between pairs of entangled particles has been consistently verified in a number of experiments of improving accuracy \cite{Aspect82, Tittel98, Rowe01}.

To implement a Bell test  for mode entanglement one must be able to measure in at least one basis other than particle number, i.e. in the superposition basis $(|0\rangle \pm |1\rangle)/\sqrt{2}$. However, the existence of any superposition of number states for massive particles is heavily debated \cite{Wick52}.  Unlike a spin state that can be rotated using an external reference frame, the magnetic field, spatial modes have no such external `knob'. Rotating the number state $|0\rangle$ to a superposed state, $(|0\rangle + |1\rangle)/\sqrt{2}$, implies the creation of a massive particle some of the time  - a physical impossibility for an isolated, cold atomic gas. This impossibility results in a \emph{superselection rule} \cite{Wick52}, stating that no coherent superposition of eigenstates of different mass can exist. 

However, recent research shows that the particle number superselection rule is not fundamental as such \cite{Bartlett07,Aharonov67, Mirman69}, but depends on a suitable reference frame \cite{Bartlett07, Dowling06}. In our case, the reference frame must play two roles.  Firstly, it should track the relative phase between the two portions of the experiment and secondly it should allow the local exchange of particles with the spatial modes. The introduction of a reference frame willprovides the tools to enable standard protocols, such as entanglement swapping, teleportation \cite{Teleport} and dense coding \cite{Heaney:09},  using spatially entangled states. It also opens the possibility to  perform a Bell-inequality test for spatial entanglement of massive bosons. 

\section{ Experimental setup and results}  

We now describe the experimental setup and calculate the expected value of the CHSH-inequality, $C$, for the simplest example of spatial entanglement, i.e. a single boson in an energy eigenstate, $|\psi \rangle$, of some confining volume, for instance, a uniform three-dimensional box or a cigar-shaped harmonic trap. The confining volume is divided into two spatial modes, $A$ for Alice and $B$ for Bob, which are individually fed into two atomic wave guides. Interpreting the number basis $\{ |0\rangle, |1 \rangle \}$ as the $z$-basis in a Bloch sphere, Alice (and Bob) can implement an effective local measurement in the $x$-$y$ plane by letting their mode impinge on a 50:50 beamsplitter where it meets a reference  state, $|\theta^{A(B)}\rangle=\frac{1}{\sqrt{2}}(|0\rangle+e^{i\theta^{A(B)}}|1\rangle)$, see Fig.~\ref{fig:setup}. The beamsplitter mixes Alice's mode, $\hat{a}$, and reference, $\hat{x}_A$, through the transformations, $\hat{c}=\frac{1}{\sqrt{2}}(\hat{a}-\hat{x}_A)$ and $\hat{d}=\frac{1}{\sqrt{2}}(\hat{a}+\hat{x}_A)$, and likewise for Bob. After the beam-splitting operation Alice (Bob) detects bosons in her (his) output ports, $\hat{c}$ and $\hat{d}$ ($\hat{C}$ and $\hat{D}$) and both select the cases when one of their two detectors registers a single particle. This implements an effective measurement of their original mode in a basis in the $x$-$y$ plane that is fixed by the phase of their individual reference state, $\theta^{A(B)}$. For instance, choosing $\theta^{A} = 0$ will project Alice's mode in the $x$-basis, $|\pm\rangle = (|0\rangle \pm |1\rangle)/\sqrt{2}$. 
To generate a second measurement setting Alice shines a laser pulse of appropriate length on her reference state, $|\theta_{1}^{A}\rangle =\frac{1}{\sqrt{2}}(|0\rangle+e^{i\theta_1^{A}}|1\rangle)$, and likewise Bob. The one-boson state $|1 \rangle$ will over time acquire an additional phase resulting in a reference state with a different phase, $|\theta_{2}^{A}\rangle$. This is comparable to altering the angle of the analyser in a conventional Bell experiment with pairs of polarised photons. 

One possible method to experimentally generate the required reference states, $|\theta^{A(B)}\rangle$, is via an atomic quantum dot (AQD) coupled to a reservoir BEC \cite{Recati05}.  For tightly confined AQDs, the large on-site interaction causes a collisional blockade, where only one or zero atoms can occupy the dot at any instance.  By controlling couplings between the AQD and the BEC, a perfectly coherent two-level state of the form $|AQD\rangle=q|0\rangle+r|1\rangle$ is realised \cite{Recati05}. If the reservoir BEC has a high average particle number, the reference state is separable with respect to the condensate and can be transferred into an atomic beamsplitter by displacing the minimum of the confining potential. The optimum choice of coefficients is $|q|=|r|=\frac{1}{\sqrt{2}}$, which will allow the maximum violation of the tested Bell inequality.  Biased values of the coefficients will downgrade the violation, but can still detect entanglement for a large range of values. 
The reservoir BEC thus plays the two specified roles.  It allows local particle exchange between the reference and the system, which enables the rotation to measurement bases other than particle number. And, it acts as a local reference that tracks the relative phase between Alice and Bob's portions of the experiment, similar to a local oscillator in a homodyne measurement \cite{Koji06}. 

Each party locally selects measurement outcomes  \cite{TomographyNote} where only one boson is detected in one of the two detectors.  The fact that a post-selection has been made, means that this is not a strict Bell inequality test.  However, {\it no} separable state (which for massive particles is $\hat\rho_{sep}=\sum_np_n|n,N-n\rangle\langle n,N-n|_{AB}$, in its most general form), will ever give rise to a violation - even with post-selection.  Our test therefore serves as a witness of mode entanglement, albeit one that illustrates that mode entanglement can be treated in an identical manner to particle entanglement.   Thus, after post-selection, Alice and Bob determine all four coincidence probabilities; for instance $ p_{cC} (\theta^A_j, \theta^B_k) = {1 \over 2} [1 + 2 \Re (\alpha \beta^*  e^{i (\theta^A_j - \theta^B_k)})]$ for Alice detecting a boson in $c$ and Bob in $C$.  Assigning the value ``+1''  for a click in detector $c\, (C)$ and ``-1'' for a click in detector $d\, (D)$ the expectation values of the joint measurements will depend on the phase of the two reference states used, $E(\theta^A_j, \theta^B_k)=\frac{1}{2}[p_{cC}(\theta^A_j, \theta^B_k)+p_{dD}(\theta^A_j, \theta^B_k)-p_{dC}(\theta^A_j, \theta^B_k)-p_{cD}(\theta^A_j, \theta^B_k)]$.  For the pure state, Eq.~(\ref{eq:purestate}),
\begin{eqnarray} 
	E(\theta^{A}_j, \theta^{B}_k) = 2 \, |\alpha| \, \sqrt{1 - |\alpha|^2}  
			\cos (\gamma + \theta^{A}_j - \theta^{B}_k),
\end{eqnarray}
where $\gamma$ is the phase of $\alpha \beta^*$ fixed by the symmetry of the populated eigenmode. $\theta^{A}_{j}$ and $\theta^{B}_{k}$ for $j, k =1,2$ are the phases of Alice or Bob's reference states. The quantity, $C$,  Eq. (\ref{eq:C}), becomes $2 \sqrt{2}$ for the optimal choice of the phases of the local measurements, $\theta^A_2=\theta^A_1+{\pi \over 2}$ and $\theta^B_2=\theta^B_1-{\pi \over 2}$ and $\theta^B_1=\gamma + \theta^A_1 + {\pi \over 4}$ and under the assumption of a symmetric split between $A$ and $B$ so that $|\alpha| = |\beta| = {1 \over \sqrt{2}}$.

\begin{figure}[t]
   \begin{center}
\includegraphics[width=0.38\textwidth]{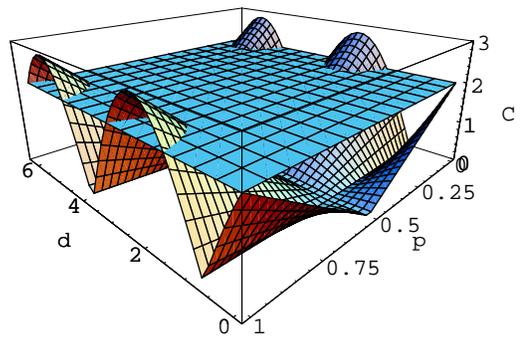}
       \caption{\label{fig:C-violation} The quantity $C$ vs. mixing probability $0 \le p \le 1$  and for phase $d = \gamma_1 + \theta^{A}_1 - \theta^{B}_1$ which can be chosen by the experimenter   to give optimal violation. Here we have assumed $|\alpha_1|^2 = |\alpha_2|^2 = |\beta_1|^2 = |\beta_2|^2 ={1 \over2}$. Violation of the inequality is achieved whenever $C > 2$. }
   \end{center}
\end{figure}

This expected behaviour of $C$ is correct under the assumption that the state is of the pure form, i.e. Eq. (\ref{eq:purestate}). However, preparing a single boson in a single energy mode cannot be achieved with perfection in an experiment. The state of the two regions will generally be mixed and can be written in the diagonalised form
\begin{equation} \label{eq:mixedstate}
	\rho_{AB} = p  | \psi_1 \rangle \langle \psi_1| + (1-p)  | \psi_2 \rangle \langle \psi_2|, 
\end{equation}
where the two states, $|\psi_1 \rangle$ and $|\psi_2 \rangle$, with coefficients $\alpha_{1,2}$ and $\beta_{1,2}$, are orthogonal and the probability of mixing is $1 \ge p \ge 0$. The theoretical prediction for the Bell-inequality quantity, $C$, is shown in Fig.~\ref{fig:C-violation} detecting entanglement for mixing probabilities below $p \approx 0.15$ and appropriate phase difference, $\theta^{A}_1 - \theta^{B}_1$. For values of $p$ outside this range, except the complete mixture at $p=\frac{1}{2}$, it was proven that there always exists a Bell-type inequality that will show a violation \cite{Horodecki95}. Moreover, when $p$ is a priori known one can use the  technique proposed in \cite{Horodecki95} to find an optimal Bell-test. Other Bell-inequalities can be tested using biased beamsplitters for which the reflectivity and transmittivity are no longer equal.

\section{ Discussion and conclusions} 

 In a Bell experiment it is essential to guarantee that no non-locality is introduced to the system during the test \cite{loopholefootnote}. In the present discussion this could potentially happen via the two reference states that are drawn from the same reservoir BEC at different locations. 
 If the reservoir BEC is, as we assume, described by a convex combination of coherent states, $\hat\rho=\frac{1}{2\pi}\int_0^{2\pi}d\theta||\alpha|e^{i\theta}\rangle\langle|\alpha|e^{i\theta}|$, i.e. it has a Poissonian particle number distribution, it is separable with respect to any spatial partitioning \cite{Simon02}. Therefore, two reference states  that are drawn from different regions of the BEC will not be entangled. If, however, the reservoir BEC has entanglement between different regions of space, which could happen when the particle number distribution is sub-Poissonian,  entanglement may emerge between the two reference states drawn from it.  In this case, entanglement between the reference states could be the origin of the violation of the Bell-inequality. However, since this is precisely the type of entanglement that we are verifying with the setup, whether it originates from the reservoir BEC or from the actual system is irrelevant for the proof that entanglement appears between spatial regions.    This contrasts with  \cite{Cooper08}, addressing single photon entanglement, where initially independent reference states become correlated after measurement.

Our results can be generalised to volumes containing $N$ bosons for which multi-dimensional Bell-inequalities exist \cite{Son06, Lee:09}. Indeed, the $N$ particle case has been studied in a different context in \cite{Laloe08}, where the violation of the CHSH inequality is assigned to the generation of spin entanglement between two condensates due to joint measurements on the condensates by two parties. However, following the results of our \paper\, one can also attribute the violation of the CHSH inequality to the spatial mode entanglement \emph{already present} in an individual condensate. By making measurements on the two condensates \cite{Laloe08}, the entanglement is swapped from between the spatial regions of the individual condensates to the spin degrees of freedom of the two condensates.

We have demonstrated that it is possible to violate a Bell-inequality with two spatial modes entangled as a result of a massive boson being coherently distributed over them.  Moreover, our scheme can be implemented by using standard atom-optics elements, such as beamsplitters and phase shifters \cite{Cassettari00}, embedded on an atomic chip \cite{Fortagh07}.  
The experimental implementation of the scheme promises the first proof of the existence of mode entanglement of fields with immediate consequences for the understanding of entanglement as an indicator of the BEC phase transitions \cite{Anders06}. 

Moreover, the implementation of our scheme is capable of proving - by contradiction - that the particle number superselection rule is not a fundamental necessity of quantum theory. Throughout our argument we assume that both, spatially entangled states and superposition states, such as the reference states exist. We then show that a Bell-inequality violation will be observed  under the assumptions made. When testing an appropriate set of Bell-inequalities for a given value of $p$ and no violation occurs then either no spatial entanglement was initially there, or the superselection rules are a fundamental feature of quantum mechanics. Conversely, an experiment implementing the test and finding a violation of the CHSH inequality  implies that the superselection rule has been overcome by coupling to a reference frame. 

\section{Acknowledgements} 

We acknowledge fruitful discussions with V. Vedral and D. Cavalcanti. J. A. is supported by the EPSRC's QIPIRC programme. L. H. is supported by EPSRC (UK) and the National Research Foundation (Singapore).


\begin{thebibliography}{99}
\bibitem{TomographyNote} Let us note that 
the full clicking pattern of the detectors builds up a tomography of the state that would, under the assumption of a correct theoretical model, allow the exact mathematical characterisation of its entanglement.
\bibitem{loopholefootnote}
Two other loopholes are the detection and the locality loophole.  
The detection loophole can be excluded when efficient single atom detection is guaranteed. Locality demands that the two arms of the setup are separated  by distances great enough to avoid communication during the measurement, which is, at present, not possible with massive particles  \cite{Rowe01}.
\end{thebibliography}
\end{document}